\documentclass{aa}
\usepackage{natbib,psfig,amsmath,amssymb}
\newcommand{\dg}{$^{\circ}$}
\def\filaa{G359.54+0.18}
\def\filab{G359.79+0.17}
\def\sagc{Sgr~C}
\newcommand{\uv}{{\it uv}}
\newcommand\kms{km~s$^{-1}$}
\begin{document}
\title{Constraints on distances to Galactic Centre
non-thermal filaments from HI absorption}
\author{Subhashis Roy}
\institute{National Centre for Radio Astrophysics (TIFR), \\
Pune University Campus, Post Bag No.3, Ganeshkhind, Pune 411
007, India.\\
E-mail: roy@ncra.tifr.res.in}

\titlerunning{HI absorption line study of three non-thermal
filaments}
\date{}

\abstract{
We have studied HI absorption towards three non-thermal
filaments (NTFs) \sagc, \filaa\ and \filab\ using the Giant
Metrewave Radio Telescope (GMRT). Our study, for the first
time, constrains the distance of the \sagc\ NTF and the HII
region seen associated with the NTF in the sky plane, to
within a few hundred parsecs from the Galactic Centre (GC).
A molecular cloud with a velocity of $-$100 \kms\ appears to
be associated with the central part of the \sagc\ NTF.  Our
study also indicates that the \sagc\ HII region is relatively
farther away than the NTF along our line of sight, and
thereby provides evidence against any possible interaction
between the two objects.  The NTF \filaa\ shows weak HI
absorption (4 $\sigma$ detection) at a velocity of $-$140
\kms, which is the velocity of a known dense molecular cloud
seen towards the NTF. This cloud is expected to be located
within $\sim$~200~pc from the GC and thereby provides a
lower limit to the distance.  The upper limit to the
distance of this NTF from the Sun is 10.5 kpc. The distance
to the NTF \filab\ is between 5.1 and 10.5 kpc from the Sun.

}

\maketitle

\section{Introduction}

The long narrow non-thermal filaments (NTFs) observed in
high resolution radio-continuum maps are unique features
seen towards only the central $\sim$~2\dg\ region of our
Galaxy.  These structures are less than 1~pc in width, but
extend up to 30~pc in length. With the exception of the NTF
named as the Pelican \citep{LANG1999.2}, which is
nearly parallel to the galactic plane, all other NTFs are
oriented perpendicular to the Galactic plane to within
20\dg\ (\citet{MORRIS1996} and the references
therein). Except the NTF \filab, which has a curved
morphology reminiscent of a partial shell, all other NTFs
portray a linear structure. The spectral indices of these
structures range from $-$0.8 to +0.3 (where S$_{\nu}
\propto \nu^{\alpha}$) \citep{ANANTHARAMAIAH1991} and they
are found to be highly polarised at centimetre wavelengths
\citep{MORRIS1996}.  Since these NTFs remain
straight despite interaction with nearby molecular clouds,
it is believed that the molecular clouds and the NTFs are in
pressure equilibrium, which indicates a magnetic field
strength of a few milliGauss inside the NTFs
\citep{YUSEF-ZADEH1987.1}.  Magnetic fields of comparable
strengths are thought to be present in the central molecular
zone (CMZ) located within $\sim$~200~pc from the centre of
the Galaxy \citep{MORRIS1996}.  Before any attempt
is made to relate the magnetic field in the NTFs with the
processes occurring in the GC, it is necessary to establish
that these NTFs are actually located in the GC region and
are not chance superpositions of foreground or background
objects \citep{LASENBY1989}.
HI absorption towards the GC `Radio-arc'
\citep{LASENBY1989} and the `Snake'
\citep{UCHIDA1992.2} have indicated that they are
located close to the GC, but the distances to the remaining
NTFs are not constrained.

An intriguing fact that has been noticed for all the well
studied NTFs is the interaction of the NTF with molecular
clouds \citep{MORRIS1996}. It appears that the
presence of an HII region near the place of interaction
plays a role in the creation and maintenance of the NTFs
\citep{SERABYN1994, STAGUHN1998,
UCHIDA1995}.
CO observations have indicated the presence of high velocity
molecular clouds; $-$65 km s$^{-1}$ and $-$130 km s$^{-1}$
towards \sagc\ \citep{LISZT1995}. It is believed
that the HII region located just south of the NTF (known as
the \sagc\ HII region) 
is actually embedded in the $-$65 km s$^{-1}$ cloud
\citep{LISZT1995, KRAMER1998}.
Two dense molecular clouds are reported to be associated
with the NTF \filaa. One of the cloud having a velocity of
$-$140 \kms, is located near the bent portion of the NTF
(`E' in Fig.~\ref{filaa.continuum}). The other cloud with a
velocity of $-$90 \kms\ is located close to the eastern edge
of the NTF \citep{STAGUHN1998}.  However, these
reported associations are based on their proximity in the sky
plane, and the spatial association of the corresponding
objects are yet to be established.

In this paper, we present new HI absorption measurements
towards three NTFs, \sagc, \filaa\ and \filab\ made with the
Giant Metrewave Radio Telescope (GMRT).  These observations
not only constrain the distances of these objects, but also
test the association of some of the above mentioned clouds
with the corresponding NTFs.  These three NTFs are located
to the south and south west of the Sgr A complex, and high
resolution radio continuum observations have confirmed the
characteristic properties of each of the NTFs: \sagc\
\citep{LISZT1995}, \filaa\
\citep{BALLY1989, YUSEF-ZADEH1997.1} and \filab\
(Lang \& Anantharamaiah, in preparation).

Due to velocity crowding near galactic longitude l=0\dg,
galactic rotation cannot be used to constrain the distances
to these NTFs.  Therefore, in this paper, detection of
absorption by known HI features has been used to provide
constraints on the distances to the NTFs. Since an
interferometer resolves the extended HI emission features, no
information about the emission features are obtained from our
observations. However, absorption by these features against
the continuum source can be observed with an interferometer.
Therefore, in this section, we briefly summarise the
distances and the velocities of the HI features identified
from single dish HI emission observations towards the NTFs
under study (e.g. \citet{COHEN1979}), which we
will refer to later in the paper.
Near the Galactic longitude of 359.5\dg, two high velocity
HI emission features known as the `Nuclear disk'
\citep{ROUGOOR1960} and the `Molecular ring'
\citep{SCOVILLE1972} have been found. `Nuclear disk'
shows high negative velocity ranging from $\approx$~$-$160
to $-$200 \kms, whereas, the `Molecular ring' has a velocity
of $\approx$~$-$135 \kms.  Both these features are believed
to be nearer than the GC and located at a distance of few
hundred parsecs from it \citep{COHEN1979}.
The emission from the `3 kpc arm'
\citep{ROUGOOR1964} located at a distance of
$\approx$~5.1 kpc from the Sun is identified at a velocity
near $-$53 \kms. At positive velocities, emission near 135
\kms\ is seen due to the HI features `XVI' and `I'
\citep{COHEN1979}, both of which are thought to be
located behind the GC. While the feature `XVI' is likely to
be located within a few hundred parsecs from the GC
\citep{COHEN1979}, the feature `I' is thought to
be 2 kpc behind the GC \citep{COHEN1975}.

In $\S$2 of this paper, we discuss the observations and data
reduction. The results from our observations are presented
in $\S$3 and their consequences have been discussed in
$\S$4. Finally, the conclusions are presented in $\S$5.

\section{Observations and data reduction}
The GMRT \citep{SWARUP1991} consists of thirty
antennas, distributed over a region of about 25 km,
with fourteen of the antennas located within a diameter of
about one km and the remaining arranged in 3 arms
each of length 14 km, shaped as an irregular Y. This
arrangement provides the necessary \uv\ coverage for mapping
both compact and extended sources.  The ratio of the longest
to the shortest baseline is $\approx$~500 with the shortest
projected baseline being $\approx$~50 metres.  Table~1 give
the details of our observations.  All the observations were
carried out in the default spectral line mode with 128
frequency channels. A bandwidth of 4 MHz was used to observe
\filaa\ and \filab\ which gave a velocity coverage of
$\pm$400 km s$^{-1}$ with a resolution of 6.7 km s$^{-1}$. The
\sagc\ complex was observed with a bandwidth of 2
MHz ($\pm$ 200 \kms) and 4 MHz ($\pm$ 400 \kms).
In all the observations, 3C286 was observed as a primary
flux calibrator, 3C287 as the bandpass calibrator and
1748-253 as the secondary calibrator.  The bandpass pattern
of the antennas change appreciably as a function of
frequency.  Therefore, we have not used frequency switching
for the bandpass calibration. Instead, we have chosen a
bandpass calibrator (3C287) with a high Galactic latitude,
so that the effect of Galactic HI absorption on its spectra
is less than 1\% \citep{DICKEY1978}. 
During our observations, the automatic measurements of
system temperatures were not implemented.  However, since in
this paper we are not concerned with the absolute value of
the flux density, we have applied an approximate correction
by scaling the flux of all the fields by the ratio of the
estimated flux density (157 mJy) of the small-diameter source
G359.87+0.18 \citep{LAZIO1999} at 1.55 GHz to the
flux density that we have estimated from our map (ratio
$\approx$~2.5). All the maps presented in this paper have
been corrected for the primary beam pattern of the antennas.

\begin{table*}·
\begin{minipage}{135mm}
\caption{Details of our observation}·
\begin{tabular}{|c c c c c c c c c c|}
\hline
Source Name & RA & Dec & l & b & Frequency  & Band- & Date & Observing & Ante- \\
        & (J2000) & (J2000) & (deg) & (deg)& (MHz) & width  &      & time (hour) &  nnas \\
         &     &     &       & &   & (MHz) &    &        &          \\
\hline
\object{\sagc} & 17 44 33.8 & $-$29 28 02 & 359.43 & $-$0.08 & 1420.4 & 2 & 2000 Oct 06 & 1 & 27 \\

\sagc\ & 17 44 33.8 & $-$29 28 02 & 359.43 & $-$0.08 & 1420.4 & 2 & 2000 Oct 08 & 2 & 27 \\

\sagc\ & 17 44 33.7 & $-$29 28 00 & 359.43 & $-$0.08 & 1420.6 & 4 & 2001 May 24 & 2.5 & 28 \\

\object{\filab} & 17 44 30.9 & $-$28 59 50 & 359.82 & 0.17 & 1420.6 & 4 & 2001 May 24 & 3.0 & 26 \\

\object{\filaa} & 17 43 48.3 & $-$29 11 26 & 359.57 & 0.20 & 1420.6 & 4 & 2001 May 25 & 3.5 & 28 \\
\hline
\end{tabular}
\end{minipage}
\end{table*}

The spectral visibility data were processed using standard
NRAO AIPS programs. Bad data due to antenna-based or
baseline-based problems were flagged using UVFLG.
The data were then calibrated and images of the fields were
formed by Fourier inversion and CLEANing (IMAGR). To reduce
the effect of comparatively large scale structures in HI
along the line of sight, we have done a high pass filtering
while making the maps of the three NTFs. Since the NTFs are
oriented almost along the east-west direction (i.e. along
{\it u}), we have filtered out data having {\it v} $\le$
2000$\lambda$ and {\it u} $\le$ 4000 $\lambda$, because of
which any 2-D structure which has a size-scale $\ge$ 1.5$'$
will not be visible and linear structures of length
$\le$ 10$'$ can only
be imaged. We have used the AIPS task UVNOU to implement
this data filtering in {\it u} \& {\it v}. The visibility data
were then used to generate the continuum and line images.
The continuum map of \sagc\ NTF made using the above
procedure is shown in Fig.~1.  Since the \sagc\ HII region
is extended, we have applied a lower \uv\ cutoff of 1
k$\lambda$ to the visibility data before making the line and
the continuum maps (Fig.~2). During observation of \filab,
the GC was located just outside the half power width of the
GMRT primary beam.  Hence, to minimise the sidelobe response
of the GC in the image of the NTF \filab, 3-D imaging (available
in the recent versions of IMAGR) was used. 

Before making the channel maps, the AIPS task UVLSF was used
to subtract a constant term across the frequency channels
corresponding to the continuum from the visibility data. The
GMRT has an FX correlator, for which `Gibbs ringing' due to
any sharp feature in the spectrum dies away much faster
($\propto$~sinc$^2$) than in a 
XF correlator. Therefore, we have not applied any spectral
smoothing to our data.  The variation in the line frequency
introduced by the earth's rotation during the observing
period is estimated to be much smaller than the frequency
channel width and so we have not applied any Doppler
corrections to the data. 
It should be noted here that the RMS noise quoted for each
spectrum is the noise estimated from the corresponding
images (i.e., from image plane) and are applicable for only 
$|$velocity$| \ge$ 20 \kms. For $|$velocity$| \le$ 20 \kms,
due to emission from the HI gas along the line of sight
(velocity crowding occurs when $|l|$ $\approx$~0\dg), the
system temperature increases and the typical RMS noise is
1.6 times higher than the quoted values. The systematic
error in our spectra (e$^{-\tau}$, where $\tau$ is the
optical depth) is believed to be less than 0.05. 

\section{Results}

In this section, we present the absorption spectra towards
the target sources and identify the velocity of the HI
absorption features. In all the spectra, unless otherwise
stated, the X-axis represents the velocity in \kms\ and
Y-axis represents the transmission (I/I$_0$), where, I is
the observed flux density of the background source at the
given frequency and I$_0$ is the actual flux density of the
source. We also assume the spin temperature of the atomic
hydrogen to be much less than the brightness temperature of
the background source and in that case I/I$_0$=e$^{-\tau}$.
All the velocities quoted in this paper are expressed with
respect to the Local Standard of Rest and the GC is assumed
to be at a distance of 8.5 kpc. As is well known, the HI
column density is related to its optical depth by the
formula, $N_H$=1.8 $\times 10^{18}\times T_S \times \int
\tau dv$, where, $N_H$ is the column density of the atomic
hydrogen, $T_S$ is the spin temperature and $\int \tau dv$
is the velocity integrated optical depth. We use this
relation to calculate the HI column density from the
observed optical depth.
\vspace{0.2cm}
\begin{figure*}
\begin{minipage}{120mm}
\hbox{
\psfig{file=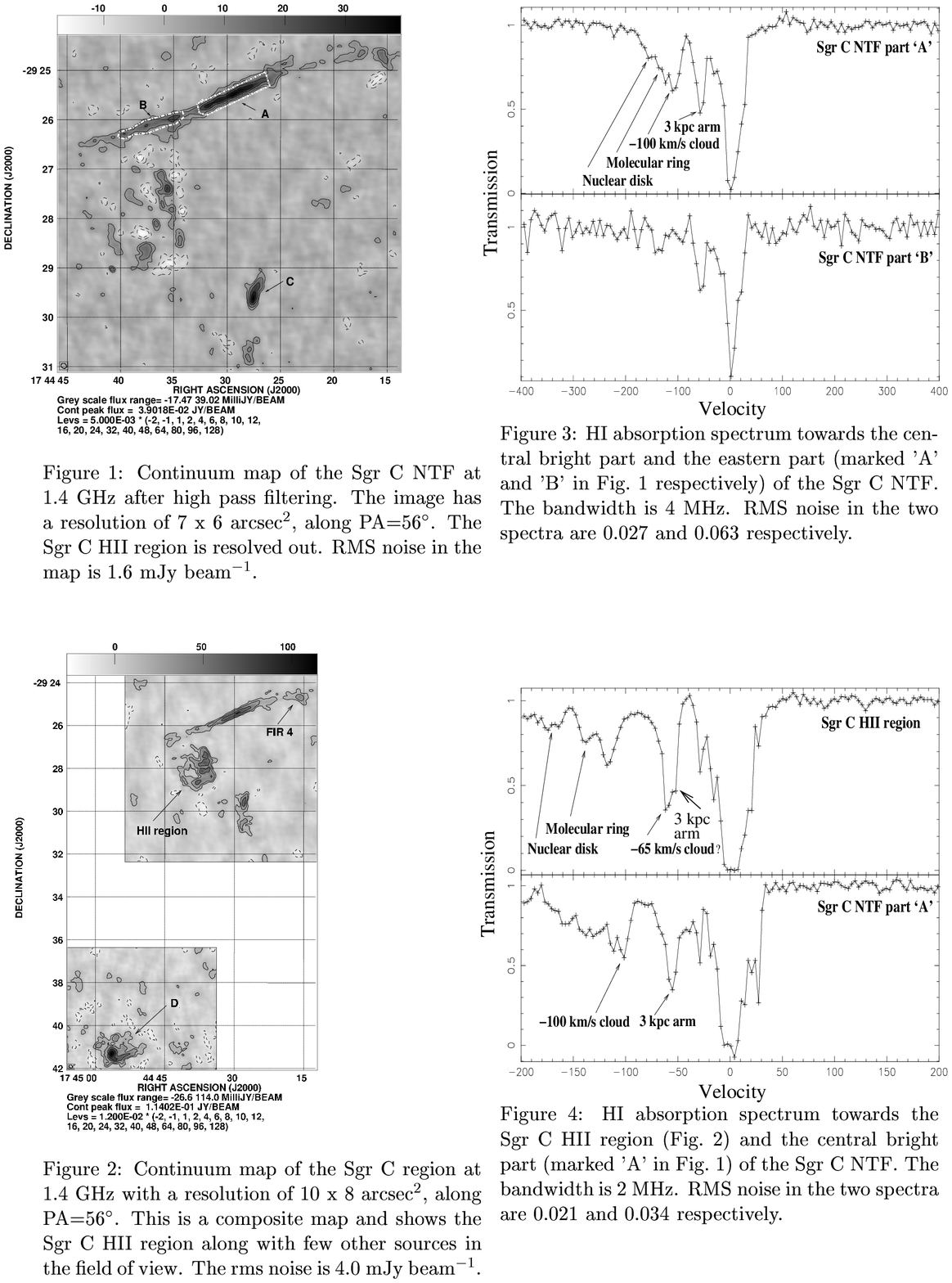,clip=,height=24.0cm}
}
\end{minipage}
\end{figure*}

\subsection{\sagc}
The absorption spectra towards various parts of the object
(Fig.~1 \& 2) are shown in Fig.~3 and 4. Fig.~3 shows the
absorption spectrum towards part `A' and `B' (see Fig.~1) of
the NTF.  The spectrum towards part `A' shows several
absorption features at negative velocities in addition to
the absorption feature at 0 \kms\ (line-width $\approx$~30
\kms).  A strong absorption feature near $-$54 \kms\ is seen
due to HI absorption by the `3 kpc arm'. A broad absorption
feature is seen between $-$100 \kms\ and $-$200 \kms, with
an almost linear decrease in optical depth from
$\approx$~0.5, at $-$100 \kms, to $\approx$~0.0, at $-$200
\kms.

The spectrum taken towards part `B' of the NTF (Fig.~3),
appears markedly different. The absorption width of the 0
\kms\ feature appears to be narrower ($\approx$~25 \kms)
than that observed towards `A'.  Absorption due to the `3
kpc arm' is also seen. However, except for a weak 4$\sigma$
absorption feature at $-$137 \kms, the wide absorption
feature between $-$100 and $-$200 \kms\ is not seen.

The absorption spectrum towards the \sagc\ HII region with a
resolution of 3.3 \kms\ is shown in Fig.~4. For comparison,
the spectrum of part `A' having the same velocity resolution
is also plotted. Both the spectra show similar absorption
features with a few differences, which we note here.
Towards the HII region, the broad absorption feature seen
between $-$100 and $-$200 \kms\ shows at least three main
components centred at $-$118 \kms, $-$138 \kms\ and $-$175
\kms\ with optical depth of $\approx$0.5, 0.3 and 0.2
respectively.  The absorption depth at these velocities are
similar to what is seen towards part `A' of the NTF.
However, the absorption feature near 0 \kms\ is broader
towards the HII region (line-width $\approx$~33 \kms).  The
HII region shows a feature at $-$61.5 \kms, in addition to
the absorption by the `3 kpc arm' observed near $-$54 \kms\
in both the spectra. However, Part `B' of the NTF being
weaker than part `A' in radio continuum, absence of the
$-$61.5 \kms\ feature could not be verified against `B'.

\subsection{HI absorption spectra towards objects located in
the field of \sagc: source C, D and FIR-4}
Fig.~\ref{sagceg.spec.4mhz} shows the spectra towards
source `C' (`3' in \citet{LISZT1995}), source `D'
and FIR-4 (Fig.~2). The spectrum towards the source `C' is
similar to what is seen towards the \sagc\ HII region, but,
in place of the wide absorption feature between $-$100 and
$-$200 \kms\ (Fig.~4), two absorption lines are observed at
$-$123 \kms\ and $-$170 \kms. Since the continuum emission
from the HII region FIR-4 is weak, its absorption spectrum
is noisy. However, we identify HI absorption at $-$135 \kms\
(5 $\sigma$ detection) and at $-$183 \kms\ (4 $\sigma$
detection) towards FIR-4.
CO emission has been observed towards the source `C' and
FIR-4 at $-$123 \kms\ and $-$135 \kms\ respectively
\citep{OKA1998.2}.
On the other hand, the spectrum towards the southern HII
region `D' (Fig.~2), (`1' in \citet{LISZT1995})
located just outside the half power width of the GMRT
antennas is markedly different.  Except the strong
absorption near the 0 \kms\ (line-width $\approx$~30 \kms), no
other absorption feature could be identified towards this
object.

\subsection{NTF \filab\ and G359.87+0.18}

The continuum image at 20 cm of the field of NTF \filab\ is
shown in Fig.~\ref{filab.continuum}. The absorption spectrum
integrated over the NTF is plotted in
Fig.~\ref{filab.spec.4mhz}. The strong absorption near the 0
\kms\ has a line-width of $\approx$~24 \kms. No absorption
feature at positive velocities is seen towards this source.
However, at negative velocities, an absorption feature can
be seen at $-$26 \kms, and a weaker feature at $-$58 \kms,
which coincides with the line of sight velocity of the `3
kpc arm'.

Fig.~\ref{filab.spec.4mhz} also shows the absorption
spectrum towards the extragalactic source G359.87+0.18.
Strong absorption near 0 \kms\ with a line-width of
$\approx$~40 \kms, and absorption at $-$53 \kms\ are
observed. \citet{LAZIO1999} have observed HI absorption
against G359.87+0.18, and the aforementioned features match
with their spectrum. However, the present observations have
a wider velocity coverage than \citet{LAZIO1999} and we
detect an additional absorption feature at +140 \kms.

\addtocounter{figure}{4}
\begin{center}
\begin{figure}
\hbox{
\psfig{file=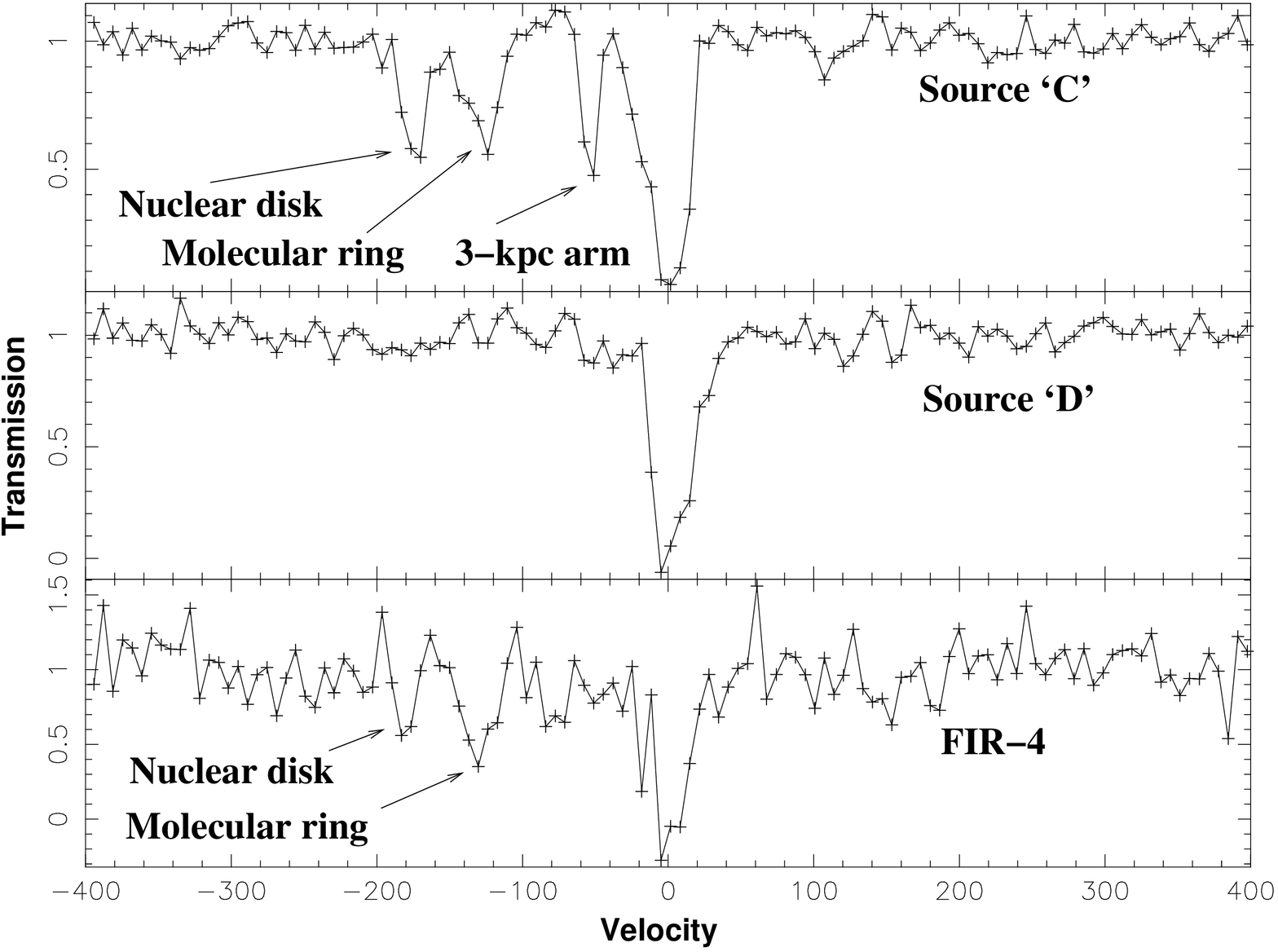,clip=,angle=270,height=6.0cm}
}
\caption{HI absorption spectrum towards the source `C'
(Fig.~1), `D' (Fig.~2) and FIR-4, located in the field of
\sagc. The bandwidth is 4 MHz.  RMS noise in the three
spectra are 0.078, 0.05 and 0.2 respectively.
}
\label{sagceg.spec.4mhz}
\end{figure}
\end{center}
\vspace{-0.6cm}

\begin{center} 
\begin{figure}
\hbox{
\psfig{file=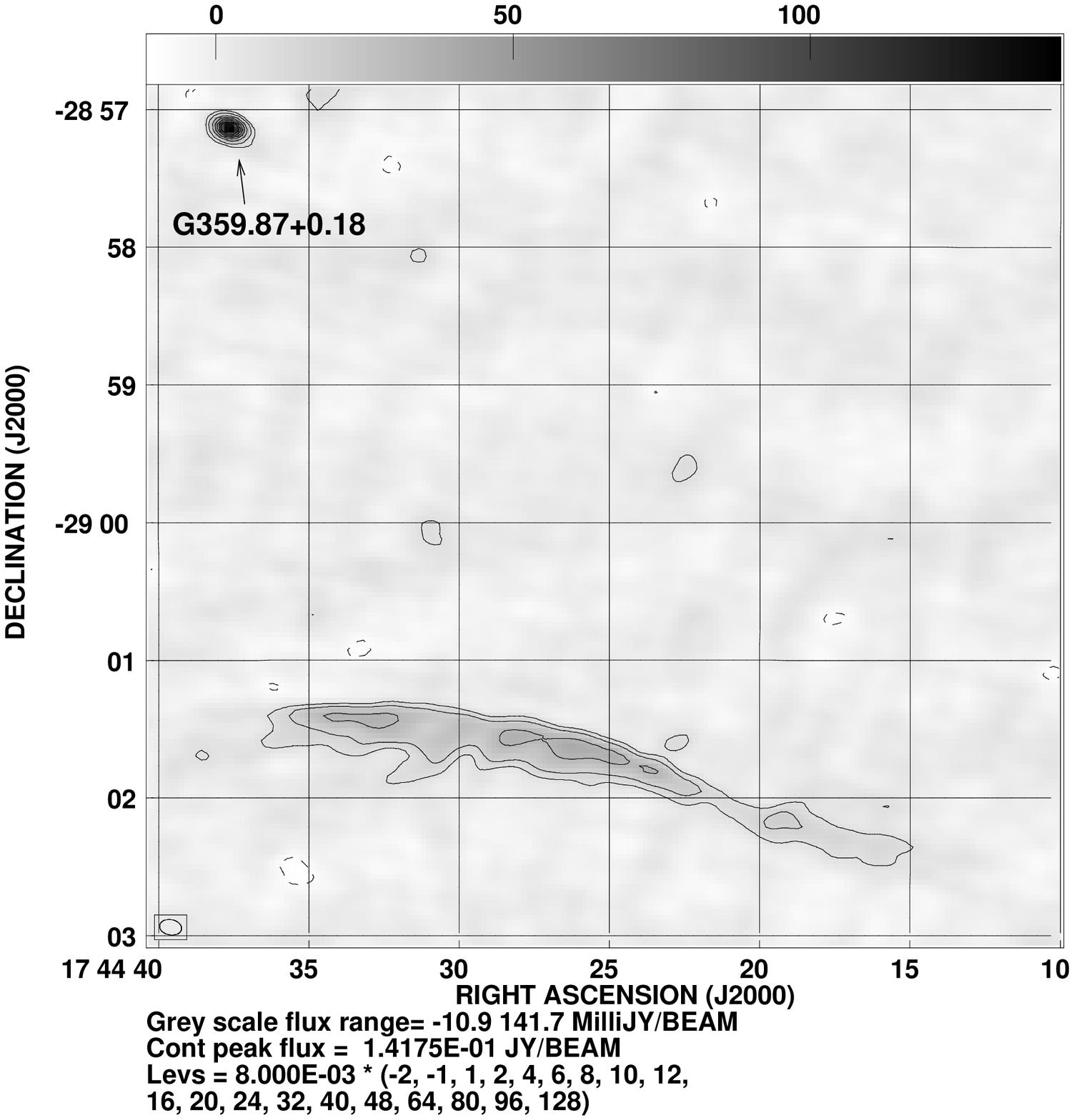,clip=,angle=0,height=9.0cm}
}
\caption{Continuum image of the NTF \filab\ at 1.4 GHz with a
resolution of 9.7 x 6.7 arcsec$^2$, along PA=79\dg. The rms
noise is 3.0 mJy beam$^{-1}$.}
\label{filab.continuum}
\end{figure}
\end{center}
\vspace{-0.6cm}

\begin{center}
\begin{figure}
\hbox{
\psfig{file=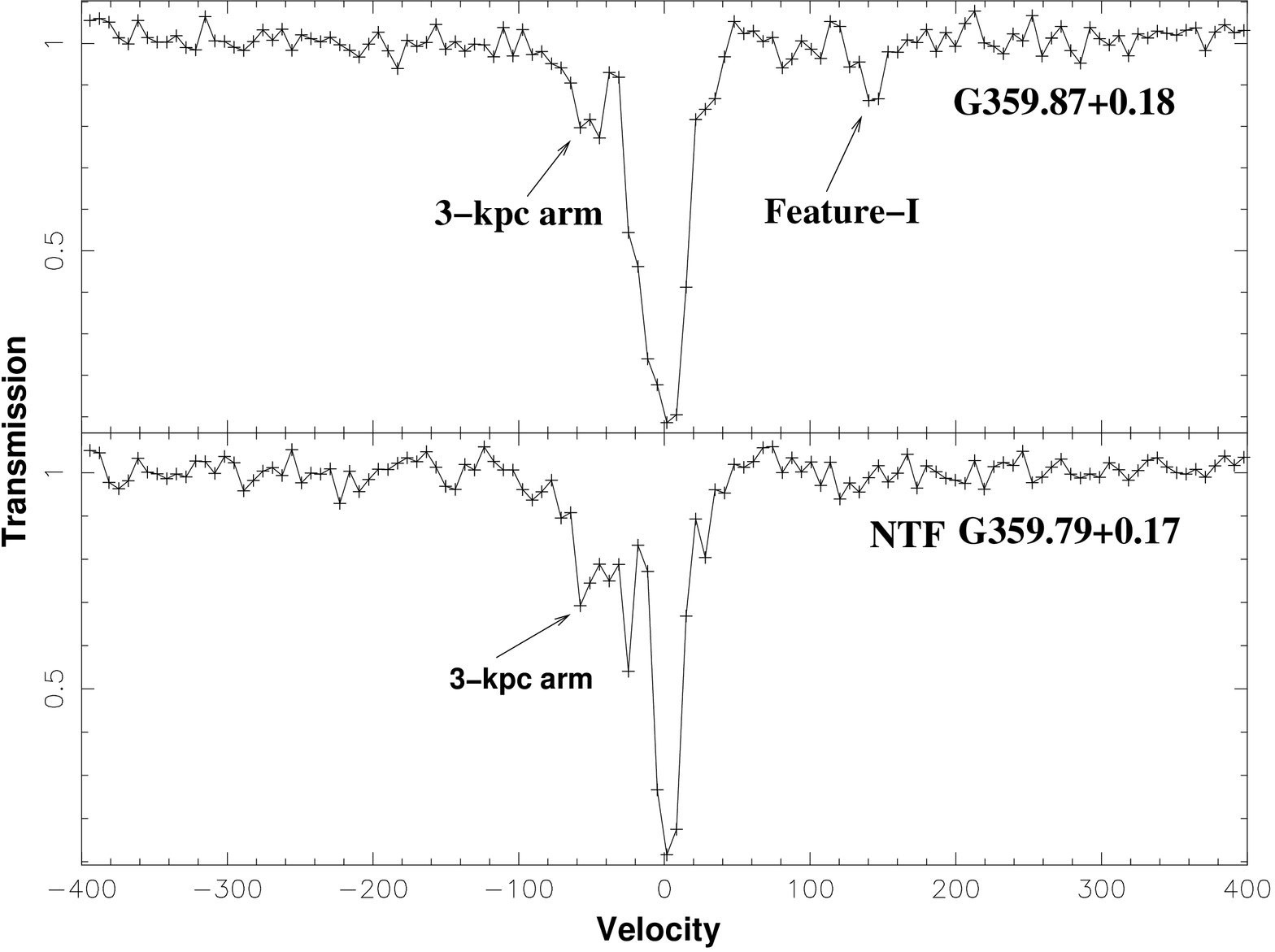,clip=,angle=270,height=5.5cm}
}
\caption{HI absorption spectrum towards the extragalactic
source G359.87+0.18 and the NTF \filab. The bandwidth is 4
MHz. RMS noise in the two spectra are 0.028 and 0.026
respectively. 
}
\label{filab.spec.4mhz}
\end{figure}
\end{center}
\vspace{-0.6cm}

%

\subsection{NTF \filaa}

\begin{center}
\begin{figure}
\hbox{
\psfig{file=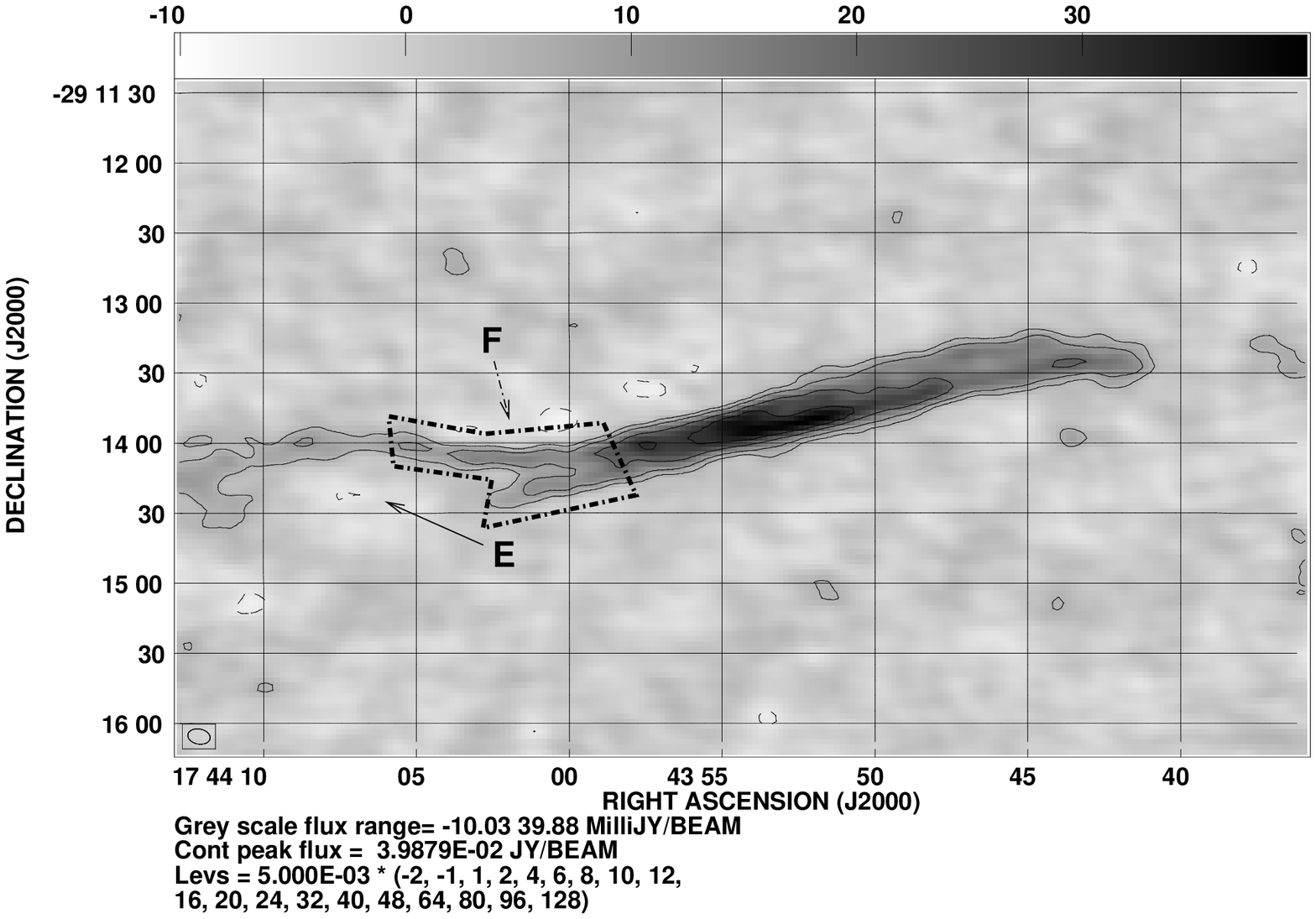,clip=,angle=270,height=6.5cm}
}
\caption{Continuum image of \filaa\ at 1.4 GHz with a
resolution of 9.6 x 6.4 arcsec$^2$, along PA=79\dg. The rms
noise is 1.7 mJy beam$^{-1}$.}
\label{filaa.continuum}
\end{figure}
\end{center}

\begin{center}
\begin{figure}
\hbox{
\psfig{file=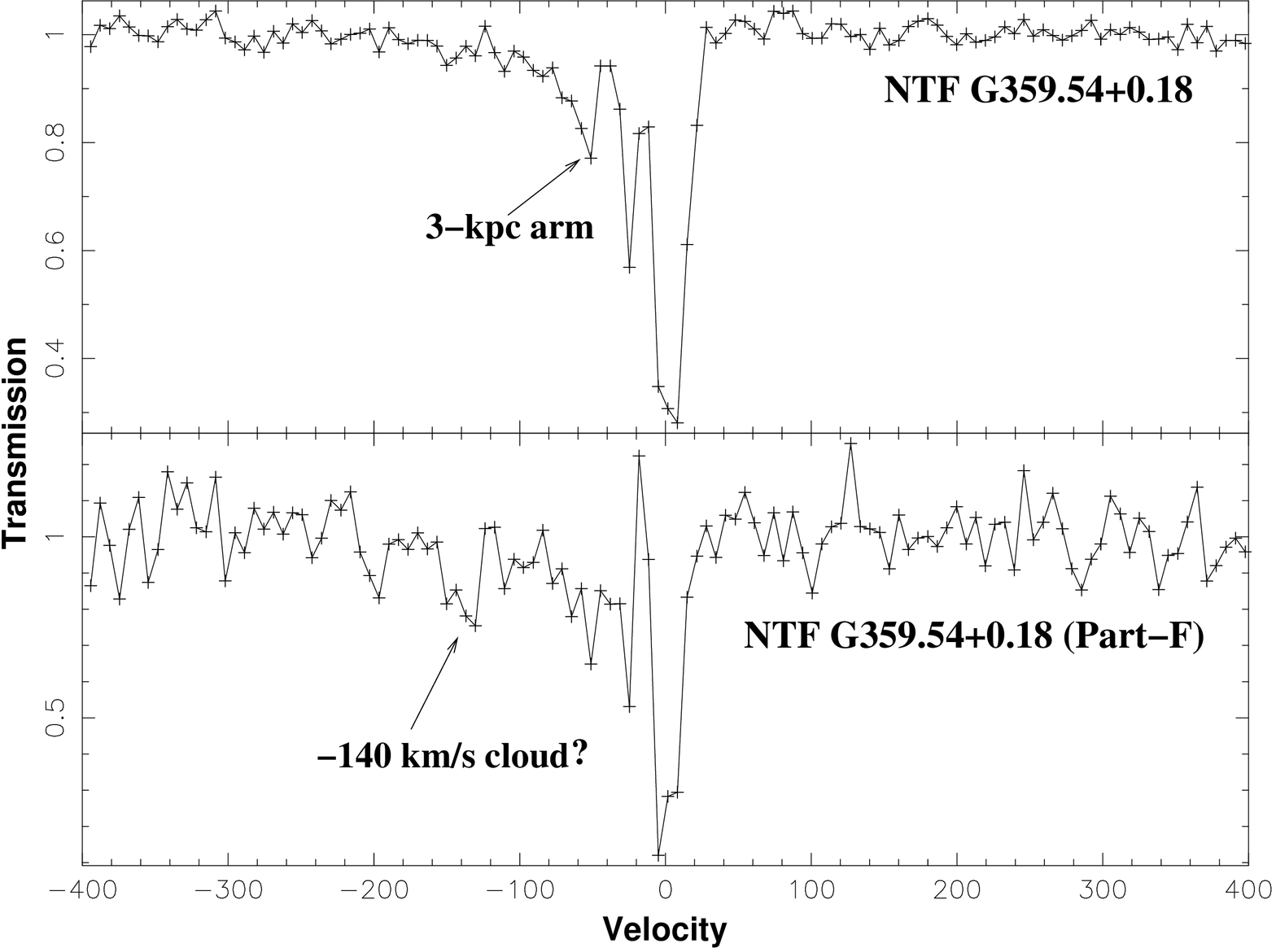,clip=,angle=270,height=6.0cm}
}
\caption{HI absorption spectrum integrated over the NTF
\filaa\ and towards a smaller portion of the NTF where
it bends (region `F' in Fig.~\ref{filaa.continuum}). The
bandwidth is 4 MHz. RMS noise in the two spectra are 0.025
and 0.1 respectively.
}
\label{filaa.spec.4mhz}
\end{figure}
\end{center}

The continuum image of the NTF \filaa\ is shown in
Fig.~\ref{filaa.continuum} and the absorption spectrum
integrated over the NTF is plotted in
Fig.~\ref{filaa.spec.4mhz}. The absorption spectrum is quite
similar to the one seen towards \filab, with components at
$-$26 \kms\ and $-$53 \kms. The strong HI absorption
near 0 \kms\ has a line-width of $\approx$~24 \kms.
\citet{STAGUHN1998} have found a dense molecular cloud at
$-$140 km s$^{-1}$ near the bent portion of the NTF
(position of the molecular cloud is denoted by `E' in
Fig.~\ref{filaa.continuum}). HI spectrum taken towards this
region of the NTF (denoted by `F' in
Fig.~\ref{filaa.continuum}) shows absorption
(Fig.~\ref{filaa.spec.4mhz}) at this velocity (4$\sigma$
detection). \citet{STAGUHN1998} also detected another
molecular cloud near $-$90 \kms\ near the eastern edge of
the NTF (not seen in Fig.~\ref{filaa.continuum}).  Since the
eastern edge of the NTF \filaa\ is faint in radio-continuum,
no useful HI absorption spectrum could be obtained.

\vspace{-0.6cm}

\vspace{-0.6cm}

\begin{center}
\begin{figure}
\hbox{
\psfig{file=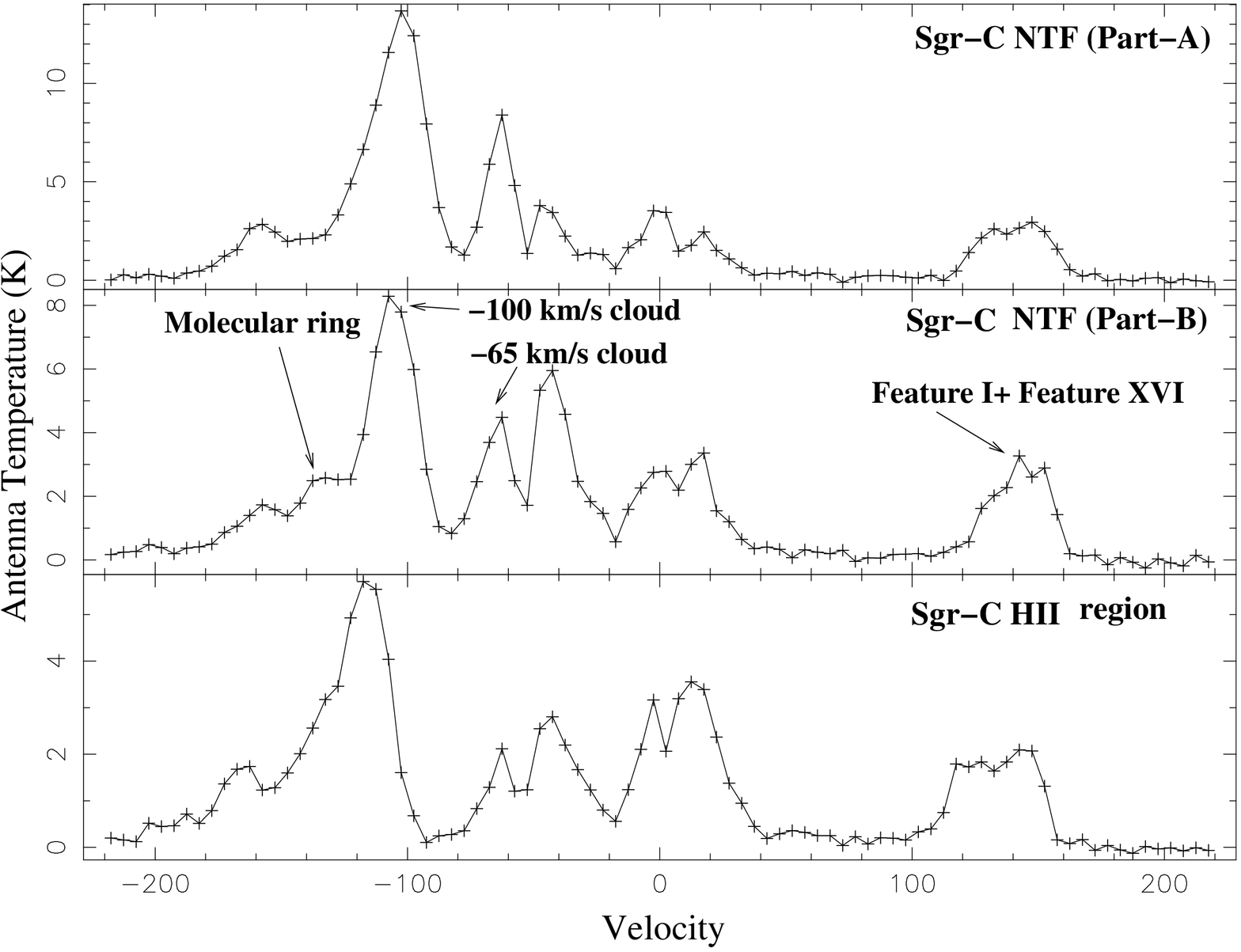,clip=,angle=270,height=6.0cm}
}
\caption{CO emission spectra towards the central part (top
panel) (marked `A' in Fig.~1) and the eastern part of the
NTF (middle), along with the spectrum taken towards the
\sagc\ HII region (bottom) (Data courtesy
\citet{OKA1998.2}).}
\label{sagc.co.spec}
\end{figure}
\end{center}
\vspace{-0.6cm}
\begin{center}
\begin{figure}
\hbox{
\psfig{file=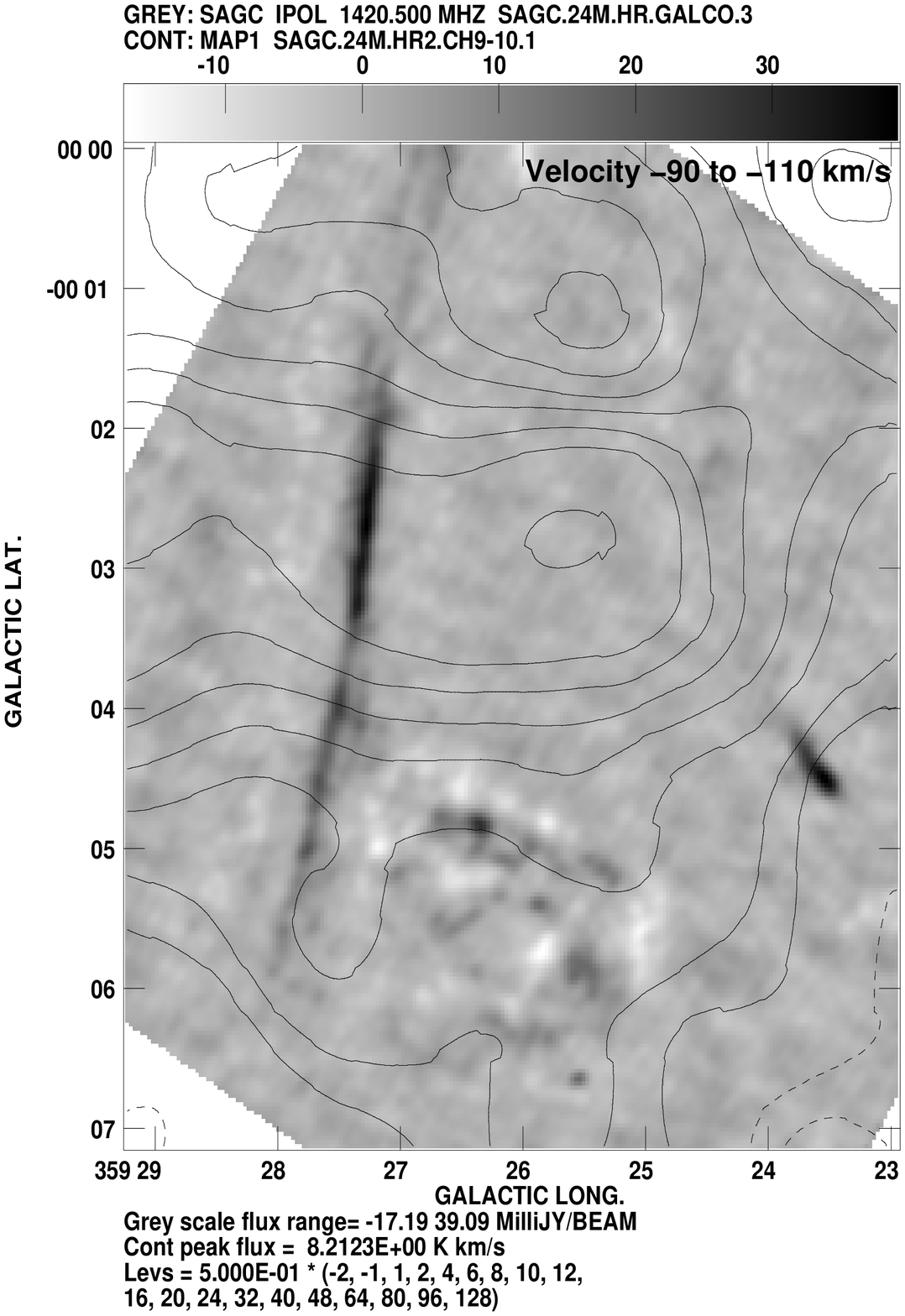,clip=,angle=0,height=9.5cm}
}
\caption{CS emission (contours) averaged over a velocity
range of $-$110 to $-$90 \kms\ superimposed on a grey scale
radio-continuum image of the \sagc\ NTF in galactic
co-ordinate. CS data courtesy \citet{TSUBOI1999}}
\label{sagc-cs-image}
\end{figure}
\end{center}
\vspace{-1.2cm}

\section{Discussion}

\subsection{Identification of HI features \& constraints on
the distances to the NTFs}

Identifications of HI absorption feature is performed by
comparison with features of known velocities.  Absorption
indicates that the continuum source is located on the far
side of the HI cloud and thereby provides a constraint on
the distance to the continuum source. The velocities and the
distances of the known HI emission features towards the
three NTFs studied here have been discussed in $\S$1 and
summarised in Table 2, which will be used to constrain the
distances to the NTFs.

\begin{table*}
\caption{Velocities and distances of the HI emission
features}·
\renewcommand{\arraystretch}{1.2}
\begin{tabular}{|l|c|c|c|}
\hline
Name of the HI & Velocity & Distance & References for \\
feature        & \kms     & kpc      & Distances \\
\hline
`0 km feature'         & $-$30 to 30 & Unknown &  \\
`3 kpc arm'             & $-$53    & 5.1 &
\citet{COHEN1976.1}* \\
`Molecular ring'   & $\approx-$ 135 & $\lesssim$ 8.5 &
\citet{COHEN1979} \\
`Nuclear disk'     & $\approx-$ 170 & $\lesssim$ 8.5 & '' \\
`XVI'              & +135           & $\gtrsim$ 8.5 & '' \\
`I'                & +135           &              10.5 &
\citet{COHEN1976.1} \\
\hline
\multicolumn{4}{l}{* Distance of `3 kpc arm' corrected for a Galactic
Centre} \\
\multicolumn{3}{l}{distance of 8.5 kpc rather than 10 kpc.} \\
\end{tabular}
\end{table*}

\subsubsection{\sagc}
Due to absorption by the line of sight HI gas and velocity
crowding near l=0\dg, strong absorption is observed near 0
\kms\ in all spectra towards the \sagc\ NTF and the HII
region discussed in $\S$3.1.  Absorption by the `3 kpc arm'
is observed at $-$54 \kms\ towards the \sagc\ NTF and the
HII region (Fig.~3 and 4).  However, the broad absorption
feature (Fig.~3) identified towards the central part of the
\sagc\ NTF (marked `A' in Fig.~1) between $-$100 and $-$200
\kms\ is peculiar.  Three distinct absorption features are
detected in the same velocity range in the higher resolution
spectrum of the HII region in Fig.~4.
This suggests absorption by clouds whose velocities differ
by an amount less than the individual line-width.

Absorption near $-$138 \kms\ is likely to be caused by the
HI associated with the `Molecular ring', which has a
line-width of $\approx$~40 \kms\ in emission
\citep{COHEN1979}. Detection of absorption beyond $-$160
\kms\ indicates absorption by the `Nuclear disk'
\citep{ROUGOOR1960}.  However, we were unable to identify
the $-$100 \kms\ absorption with any known HI emission
feature.
We believe that this feature was missed due to beam dilution
in the low resolution (single dish) HI surveys.
%
We examined the existing CO \citep{OKA1998.2} and CS
\citep{TSUBOI1999} emission line maps of this region
(spatial resolution $\sim 1'$) made using the 45 m telescope
of the Nobeyama Radio Observatory. Since these spectra have
a much higher angular resolution than the single
dish HI surveys, we attempted an identification of the
$-$100 \kms\ feature with a CO or CS feature. The CO spectra
towards part `A' and `B' of the NTF and the HII region is
shown in Fig.~\ref{sagc.co.spec}. Strong CO emission is
indeed observed near $-$100 \kms\ (line-width $\approx$~20
\kms).  The antenna temperature from part `A' of
the NTF is a factor of 2 higher than part `B'.  The CS
spectrum also shows a similar ratio. (CS emission traces
dense molecular clouds with density $\sim 10^4$ cm$^{-3}$,
which are typically found in the CMZ).
Fig.~\ref{sagc-cs-image} shows a contour image of this cloud
in CS emission (data courtesy \citet{TSUBOI1999}) within a
velocity range of $-$110 and $-$90 \kms.  Note that the
molecular cloud covers almost the whole NTF, although the
$-$100 \kms\ absorption is only observed towards part `A'.

%
%
%
The $-$100 \kms\ molecular cloud has been catalogued
by \citet{OKA2001.2} and by
\citet{MIYAZAKI2000} (cloud 17 at l=359.48\dg,
b=$-$0.042\dg). Using the estimated parameters of the cloud
(radius $\approx$~5~pc, mass 7.2$\times 10^4 M_{\odot}$),
the mean density of this cloud is 3100 H$_2$ cm$^{-3}$.
Assuming the ratio of atomic to molecular hydrogen is 0.01
\citep{LISZT1983, LASENBY1989}, the HI column density is
9.3$\times$10$^{20}$ cm$^{-2}$. We find that an assumed spin
temperature of 100 K, and a line width of 20 \kms\ explains
the observed HI optical depth of $\approx$~0.5. Hence, we
believe that we are observing absorption by HI associated
with the $-$100 \kms\ molecular cloud.

Our high resolution spectrum of the \sagc\ HII region
clearly shows the presence of two spectral features in the
range of $-$50 to $-$65 \kms (Fig.~4). \citet{LISZT1995}
have identified a molecular cloud towards \sagc\ with a
velocity of $\approx$~$-$65 \kms, which they call
M359.5$-$0.15. The $-$54 \kms\ feature is identified with
the 3 kpc arm, whereas the  $-$61.5 \kms\ feature in our HI
data is believed to arise from M359.5$-$0.15.
The presence of absorption by the molecular cloud
M359.5$-$0.15 indicates that the \sagc\ HII region is either
embedded in or located at the far side of the cloud. This
result supports the suggestion of \citet{LISZT1995} that
\sagc\ HII region (Fig.~2) is located in a cavity of
M359.5$-$0.15.  We note that CO emission near $-$65 \kms\
has been detected towards both part `A' of the NTF and the
HII region (Fig.~10). The lack of HI absorption towards part
`A' of the NTF indicates that it is located in the front of
the $-$65 \kms\ cloud.  Therefore, we believe that the
\sagc\ HII region is relatively farther away than the \sagc\
NTF, which provides evidence against any interaction between
the two (see the schematic in Fig.~13).
%

Having identified the absorption features in the HI
spectrum, here we estimate the distance of the \sagc\
complex. Since absorption by the `Molecular ring' and the
`Nuclear disk' have been detected towards the \sagc\ NTF
(part `A' in Fig.~1) and the HII region, we can conclude
that these objects seen in radio continuum are located at a
minimum distance of these HI features.  As `Molecular ring'
and the `Nuclear disk' are located within $\sim$~200~pc from
the GC, this provides a lower limit of $\approx$8.3 kpc to
these objects.
We note that despite the emission feature seen in the CO map
at $\sim$140 \kms\ (Fig.~\ref{sagc.co.spec}), no
corresponding HI absorption could be detected towards the
\sagc\ region.  The CO emission from the molecular cloud at
$\sim$140 \kms\ is likely to be associated with the HI
features `XVI' and `I' and both these features are located
at the far side of the GC. Absence of any absorption by the
HI associated with these structures indicate that the \sagc\
NTF and the HII region are located within a few hundred
parsecs at the far side of the GC, which provides an upper
limit to their distances.

There appears to be weak absorption (4 $\sigma$ detection)
towards part `B' of the NTF at $-$137 \kms\, which appears
to have been caused by the `Molecular ring'. Since we expect
a much stronger absorption if this part of the NTF is
located at the far side of the `Molecular ring', we suggest
that the part `B' of the NTF is embedded in the `Molecular
ring'. Also, since the `Nuclear disk' is expected to be
located farther away from the GC than the `Molecular ring'; lack
of absorption by this feature may indicate that there is a
hole in the `Nuclear disk' towards part `B' of the NTF.

\subsubsection{Objects located in the field of \sagc: source
C, D and FIR-4}
The HI spectrum ($\S$3.2) towards source `C' (Fig.~1) shows
absorption at $-$53 \kms, $-$123 \kms\ and $-$170 \kms\
(Fig.~\ref{sagceg.spec.4mhz}), which indicates HI absorption
by the `3 kpc arm', `Molecular ring' (`Molecular ring' is
expected to have a velocity of $\approx$~$-$135 \kms, which
differs in this case by less than the line-width) and the
`Nuclear disk' respectively.  HI spectrum of FIR-4 also
indicates absorption by the `Molecular ring' at $-$135 \kms\
and the `Nuclear disk' at $\approx$~$-$183 \kms\
(Fig.~\ref{sagceg.spec.4mhz}).
Since the `Molecular ring' and the `Nuclear disk' are
thought to be located within a few hundred parsecs from the
GC, HI absorption by these two features provides a lower
limit of $\approx$~8.5 kpc to the distances of the source
`C' and FIR-4.  On the other hand, the spectrum taken
towards the southern HII region `D' (Fig.~2), shows no
absorption near $-$53 \kms\ (Fig.~\ref{sagceg.spec.4mhz}).
\citet{LISZT1995} have suggested that source `D' is a
local object due to the absence of non-zero velocity
features in the HI spectrum. The absence of any non-zero
velocity features in our spectrum also makes us to draw a
similar conclusion.

\subsubsection{NTF \filab\ \& G359.87+0.18}

HI absorption near +140 \kms\ is observed
(Fig.~\ref{filab.spec.4mhz}) towards the extragalactic
source G359.87+0.18.  The HI emission feature `I' seen in HI
\citep{COHEN1979} is located $\approx$~2 kpc
behind the GC and has a line of sight velocity of +135 \kms\
at this longitude. The HI feature `XVI' located a few
hundred parsecs behind the GC also has a similar velocity at
this longitude. The absorption seen in
Fig.~\ref{filab.spec.4mhz} near +140 \kms\ matches closely
with the velocity of these two features, which indicates
that the absorption is caused by either one or a combination
of both these features.

\begin{center}
\begin{figure}
\hbox{
\psfig{file=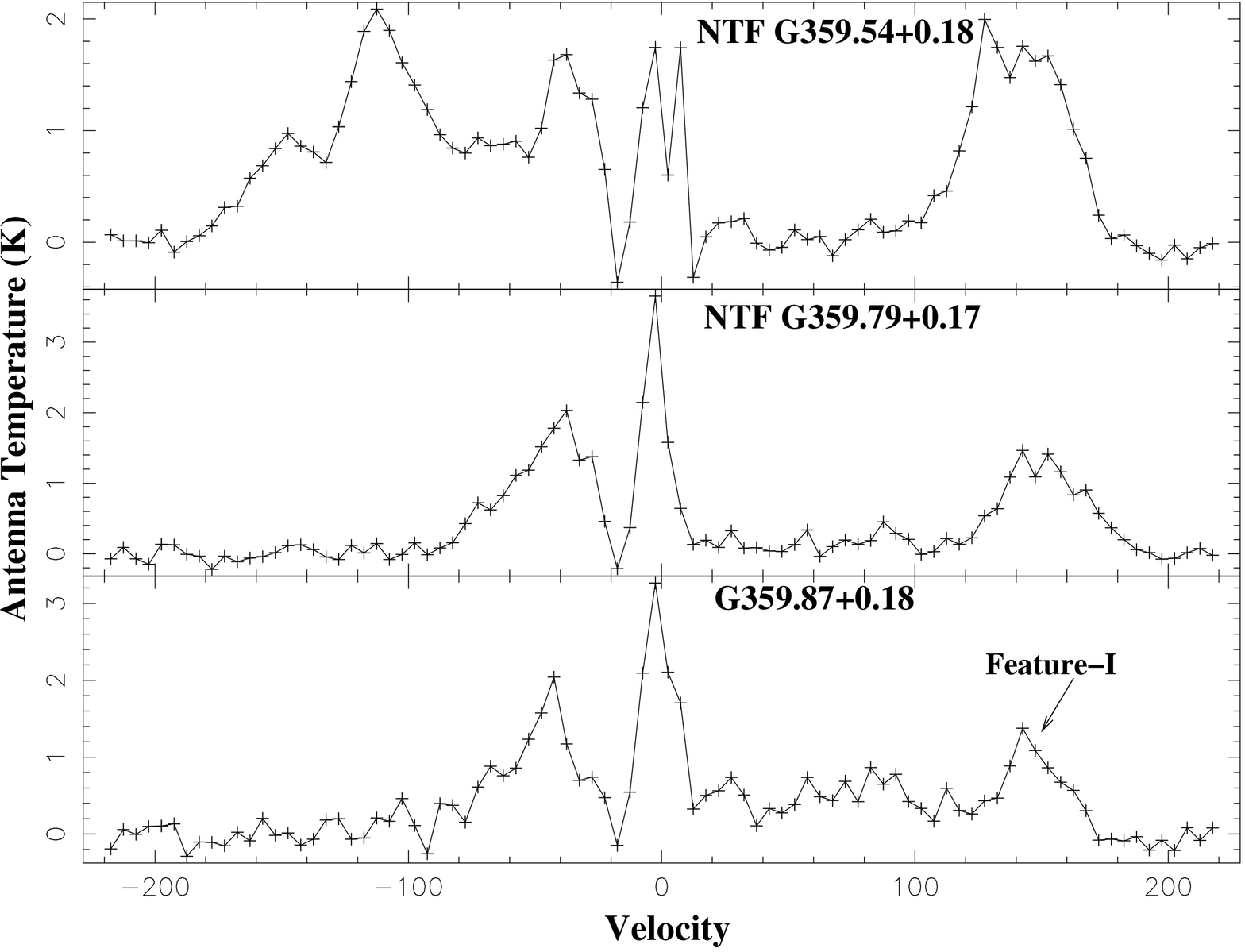,clip=,angle=270,height=5.5cm}
}
\caption{CO emission spectrum towards the NTF \filaa\
(top), \filab\ (middle) and the extragalactic source
G359.87+0.18 (bottom). Data courtesy \citet{OKA1998.2}}
\label{fila1.fila2.eg.co.spec}
\end{figure}
\end{center}

No HI absorption at high positive velocity is detected
towards the NTF \filab.  However, CO emission has been
detected near +140 \kms\ towards both \filab\ and the
extragalactic source G359.87+0.18
(Fig.~\ref{fila1.fila2.eg.co.spec}), which indicates that
there is no hole in feature `I' (or perhaps in feature
`XVI') along these directions.  Consequently, the upper
limit to the distance of the NTF is $\approx$~10.5 kpc. The
presence of absorption in the spectrum of the NTF up to a
negative velocity of $-$58 \kms\ suggests absorption by the
`3 kpc arm' and consequently, the lower limit to its
distance is $\approx$~5.1 kpc from the Sun.

\citet{LAZIO1999} suggested the presence of a $-$20
\kms\ cloud at the far side of the GC.  The presence of a
narrow absorption feature $-$26~\kms\ in our data towards
the NTF \filab\ may indicate that it is embedded in this
negative velocity cloud.  This cloud could have caused the
somewhat wider absorption profile in the spectrum towards
the extragalactic source G359.87+0.18 which is observed
through a substantially longer line of sight path through
this HI gas.

\vspace{-0.6cm}

\subsubsection{NTF \filaa}

The spectrum of \filaa\ (Fig.~\ref{filaa.spec.4mhz}) is
quite similar to that towards the NTF \filab. The presence
of absorption by the `3 kpc arm' near a velocity of $-$53
\kms\ clearly indicates that \filaa\ is located
beyond 5.1 kpc from the Sun. As \filab\ and G359.87+0.18,
CO emission has also been detected towards this NTF at
140 \kms\ (Fig.~\ref{fila1.fila2.eg.co.spec}),
but, absence of HI absorption at this high positive velocity
indicates that it is located within $\approx$~10.5 kpc from
the Sun.  As described in section $\S$4.1.3, this NTF also
seems to be embedded in the $-$20 \kms\ cloud. We could
detect weak HI absorption at 4$\sigma$ level at 
$-$140 \kms\ (Fig.~\ref{filaa.spec.4mhz}) toward part-F of
the NTF shown in Fig.~\ref{filaa.continuum}. A dense
molecular cloud having the same velocity is also found to be
present near this location \citep{STAGUHN1998}. Since dense
molecular cloud are typically found only in the CMZ, our
observations suggest that the NTF \filaa\ is also embedded
in or located at the far side of the CMZ.

\begin{table*}[!ht]
\begin{minipage}{200mm}
\caption{HI column densities toward \sagc, \filab, \filaa\
and G359.87+0.18}
\baselineskip 20 pt
\begin{tabular}{|l c c c c c|}
\hline
Column density & Central part of & \sagc\ HII & \filab\ & \filaa\ & G359.87+0.18 \\
 (cm$^{-2}$)   & \sagc\ NTF & region     &         &         &      \\
               &                 &            &         &         &          \\
\hline
N(H) (+155 to +125 \kms)&     -- & --         & --      & --      &  5.2 $\times$ 10$^{18}$ Ts  \\

N(H) (+35 to $-$35 \kms) & 1.6 $\times$ 10$^{20}$ Ts  & 2.3 $\times$ 10$^{20}$ Ts & 8.8 $\times$ 10$^{19}$ Ts & 6.3 $\times$ 
10$^{19}$ Ts & 1.3 $\times$ 10$^{20}$ Ts \\

N(H) ($-$77 to $-$40 \kms)  & 3.0 $\times$ 10$^{19}$ Ts & 2.7 $\times$ 10$^{19}$ Ts & 1.8 $\times$ 10$^{19}$ Ts & 1.0 $\times$ 10$^{19}$ Ts & 1.9 $\times$ 10$^{19}$ Ts \\

N(H) ($-$90 to $-$200 \kms) & 4.8 $\times$ 10$^{19}$ Ts & 3.6 $\times$ 10$^{19}$ Ts & -- & -- & -- \\

\hline
\end{tabular}
\end{minipage}
\end{table*}

\subsection{HI column densities}
We have estimated column densities for the HI feature `I'
(+155 to +125 \kms), the local foreground HI component (+35
to $-$35 \kms), the `3 kpc arm' ($-$77 to $-$40 \kms), and
the combined molecular cloud and `Nuclear Disk' ($-$90 to
$-$200 \kms) components towards the NTFs \sagc, \filab\ and
\filaa\ and the extragalactic source G359.87+0.18.  These
column densities, expressed as a multiple of the spin
temperature `Ts', are presented in Table~3. The velocity
limits for the various HI features were determined by visual
inspection of the absorption spectra towards these sources.
Since we could not separate the absorption by the molecular
cloud and the `Nuclear disk', we have quoted their total
column densities.  Due to the velocity crowding towards the
GC, the optical depth of the local line of sight HI gas is
large (i.e., $\exp (-\tau) \approx$~0).  Therefore, any
small error in estimating the optical depth would result in
comparatively larger error in the estimated column
densities. Hence, the estimated total column densities of
the foreground HI gas should be taken with caution.  The
column density of the line of sight HI gas ($-$35 \kms\ $<$
velocity $<$ 35 \kms) shown in Table~3, falls by a factor of
2 towards the NTFs \filab\ and \filaa\ for $|$b$|$
$\approx$~0.2\dg, which suggests that the scale height of
the HI gas at the distance of the GC is $\sim$~27 pc.

The HI column density estimated from the optical depth
towards the `Snake' (G359.1$-$0.2) NTF
\citep{UCHIDA1992.2} is 1.1 Ts $\times$ 10$^{20}$ due
to the local foreground HI component, 2.3 Ts $\times$
10$^{19}$ due to the `3 kpc arm', and 2.8 Ts $\times$
10$^{19}$ due to the `Molecular ring'. These numbers are
comparable (within a factor of two) with our estimated
column density towards \filaa\ and \filab\ given in Table~3.
We note that the column density as quoted in Table~1 of
\citet{UCHIDA1992.2} towards all of their sources are
incorrect and 100 times higher than what is found from their
quoted optical depth.

\subsection{Interaction of the molecular cloud with the
\sagc\ NTF}

As shown in $\S$3.1, no HI absorption by the $-$100 \kms\
cloud could be identified towards part `B' of the \sagc\ NTF
(Fig.~3). However, HI absorption by the $-$100 \kms\ cloud
could be identified towards part `A' of the \sagc\ NTF. From
the ratio of CO brightness, temperature towards part `A' and
`B' of the NTF ($\S$4.1.1), we expect HI optical depth of
$\approx$0.2 towards `B' in the velocity range of $-$95 and
$-$125 \kms, which is 5 times higher than the effective
noise in the spectrum.
Therefore, if the abundance of atomic hydrogen and the spin
temperature (Ts) are the same in different parts of the
cloud, HI absorption towards part `A' and lack of it towards
part `B' of the NTF can only be explained if `A' is located
at the far side and `B' is located at the near side of the
cloud and the NTF is at least partly embedded within the
cloud. In Fig.~13, we show a schematic of the \sagc\
complex in the GC region, including the relative location of
the absorbing clouds about the \sagc\ complex.

\begin{figure*}
\hbox{
\psfig{file=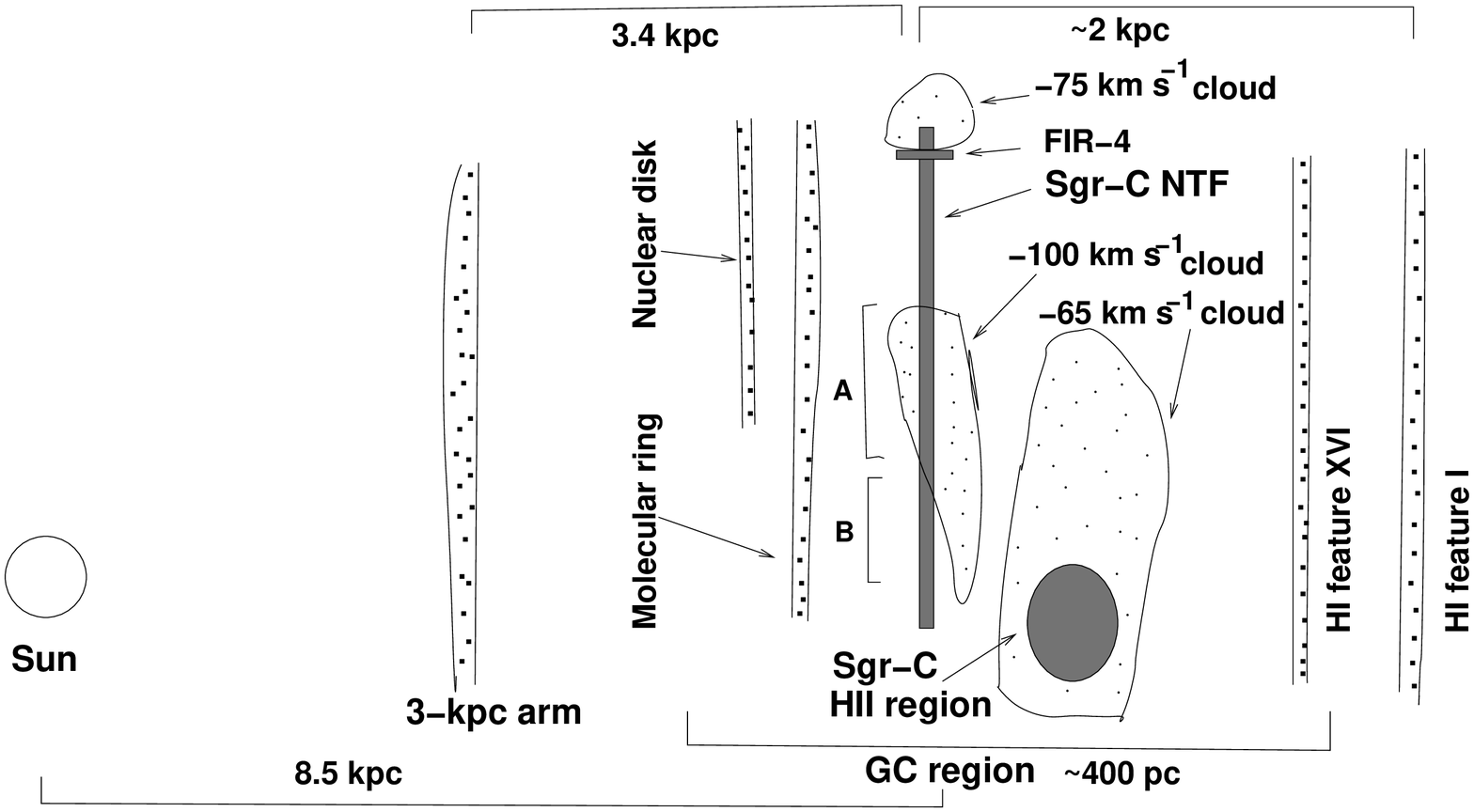,clip=,angle=0,width=16.0cm}
}
\caption{Schematic diagram of the \sagc\ complex with
HI absorbing clouds (not to scale) as seen from
bottom of the Galaxy}
\label{sagc.figure}
\end{figure*}

Here, we consider three possible scenarios, which could have
led the \sagc\ NTF to be partly embedded within the $-$100
\kms\ cloud.
\newline
(i) We first consider the case of the NTF tunnelling
through the already existing $-$100 \kms\ cloud after
being created. To penetrate the cloud, the high energy
electrons in the NTF have to first ionise the neutral gas.
With an estimated width of 35$^{''}$ (1.4 pc), length
255$^{''}$ (10~pc) and equipartition magnetic field of 100
$\mu$G \citep{ANANTHARAMAIAH1991} in the NTF, the estimated
total energy of the relativistic electrons is
$\sim$2~$\times$~10$^{47}$ erg. However, the energy required
to ionise even a HI cloud (mean density 3100 cm$^{-3}$) of length
equal to that of the size of the cloud (10 pc) and
cross-sectional area equal to that of the NTF is $\sim$3
$\times$ 10$^{49}$ erg. Therefore, the energy of the
relativistic electrons is less than 1\% of the required
energy to penetrate the molecular cloud. Hence, unless a
tunnel in the cloud already exists, the electrons will not
have sufficient energy to support this scenario.
\newline (ii) Next, we consider the case, where the NTF does
not penetrate the cloud, but the synchrotron electrons are
generated from the ionised surfaces on either side of the
$-$100 \kms\ cloud and follow the local magnetic field
lines. In the case of `Radio-arc', which is comprised of
several narrow filaments, the individual filaments appear to
either end or begin from the HII region G0.18$-$0.04 (sickle
like feature), located near the midpoint of the NTF.  Based
on the observations of the `Radio-arc', \citet{SERABYN1994}
have proposed that association of the NTF, molecular cloud
and HII region are necessary to explain the generation of
high energy electrons forming the NTFs. In Fig. 2c of
\citet{LISZT1995}, we observe two filamentary structures
towards the western side of the part `A' of \sagc\ NTF,
which merge into a single brighter filament near `A'. 
Our observations indicate that the $-$100 \kms\ cloud is
associated with the NTF near `A' (RA (J2000)=17h44m33s,
DEC=-29\dg25'55$^{''}$). In this scenario, if the surface of
this cloud is ionised, then association of the molecular
cloud and HII region with the NTF satisfies the criterion of
\citet{SERABYN1994} to generate it. Here, if the electrons
flowing towards part `A' is at the far side and part `B'
lies at the near side of the cloud, then the absence of
$-$100 \kms\ HI absorption against part `B' can be
explained. However, no such ionised surface of the cloud (as
HII region) like G0.18$-$0.04 has been detected near part
`A' of the NTF from the Nobeyama millimetre array
observation of this region \citep{TSUBOI1991} at 22 GHz.
Thus, the above scenario also fails to explain the
observations. \newline
(iii) Finally, we consider an alternative source of the
synchrotron electrons in the \sagc\ NTF and the case of
collision of the $-$100 \kms\ cloud with this already
existing NTF. \citet{LISZT1995} proposed that the HII region
FIR~4 \citep{ODENWALD1984} is associated with the \sagc\
NTF, which acts as the source of relativistic electrons.
Fig.~7 in \citet{LISZT1995} shows a molecular cloud around
this HII region. We note that FIR-4, with its wedge like
morphology, is similar to the HII region G0.18$-$0.04,
discussed earlier.
The longer side of FIR-4 lies almost perpendicular to the
direction of the NTF, thereby suggesting interaction between
the NTF and the HII region in a way similar to the Radio-arc.
The CS data \citep{TSUBOI1999} shows a compact dense
molecular cloud with a velocity of $-$75 \kms, which
coincides with the position of FIR-4.  However, in the
absence of any recombination line detection from this HII
region, actual association of FIR-4 and the $-$75 \kms\
cloud cannot be verified. In $\S$4.1.2, it is shown that the
lower limit to its distance is $\approx$~8.5 kpc, which is
consistent with its proposed association with the NTF.
Therefore, if we assume that the FIR-4 acts as the source of
relativistic electrons to the \sagc\ NTF, collision of the
$-$100 \kms\ cloud with the central part of
the NTF, which causes the NTF to appear partly embedded in
the cloud can also explain the observations. After the
collision of the cloud with the NTF, if the magnetic
pressure in the NTF would have been less than that of the
cloud, the fields could be pinched or disrupted and the flow
of high energy electrons towards part `B' significantly
reduced. On the other hand, if the magnetic pressure in the
NTF is much larger than the pressure due to the cloud
\citep{YUSEF-ZADEH1987.1}, then the NTF will remain stable
and the cloud may wrap around the NTF.  Here, the flow of
electrons from the FIR-4 to part `B' of the NTF will not be
significantly disrupted. 

If flow of electrons has stopped in part `B' (first
possibility), then this part of the NTF is likely to show a
steeper spectral index.  However, the estimated spectral
index between 1.6 GHz and 330 MHz is quite flat, which
\citet{LAROSA2000} has suggested to be due to the presence
of thermal emission from the nearby HII region. In this
regard, we note that if the $-$100 \kms\ cloud is moving
with a velocity similar to its line of sight velocity in the
sky plane, then the collision took place around
5$\times$10$^4$ years ago.  However, to our knowledge, part
`B' of the NTF has been imaged with high resolution and
sensitivity only up to a few GHz and at this frequency range,
the half-life of synchrotron electrons responsible for the
emission in an equipartition field of 0.1 mG
\citep{LAROSA2000} is $\sim$~10$^5$ years.  Therefore,
unless the magnetic field is much higher than the
equipartition value, no significant steepening of the
spectral index may be detectable in the presently available
data. Hence, to distinguish between the two cases, future
polarimetric observations, which traces the direction of the
local magnetic field lines, and can show that whether the
magnetic field lines in part `B' are pinched, will be
useful. We prefer the third model, taking into consideration
the available information. Since, the NTF does not show any
change of shape or direction near the place of interaction
with the $-$100 \kms\ cloud, the magnetic pressure in the
NTF is likely to be more than the pressure of the cloud.

\section{Conclusions}
HI absorption studies of three NTFs known as the
\sagc, \filaa\ and \filab\ using the GMRT have yielded the
following results: \\
\indent
(a) For the first time, the \sagc\ NTF and the HII region
are shown to be located within a few hundred parsecs from
the GC.

(b) Our study indicates that the \sagc\ HII region is either
embedded in or located behind the $-$65 \kms\ molecular
cloud, whereas the \sagc\ NTF is located at the near side of
the cloud, which argues against any possible interaction
between the two objects.

(c) A molecular cloud with a velocity of $-$100 \kms\
appears to be associated with the central part of the
\sagc\ NTF, and on the basis of the presently existing data,
it appears that the magnetic pressure in the NTF is higher
than the pressure due to the $-$100 \kms\ cloud.

(d) HI absorption by the `3 kpc arm' is detected against all
the three NTFs, which indicates that the NTF \filaa\ and
\filab\ are located at a minimum distance of 5.1 kpc from
the Sun.

(e) Weak HI absorption (4 $\sigma$ level) at $-$140 \kms\
suggests that the NTF \filaa\ is located at a minimum
distance of $\approx$~8.5 kpc from us.

(f) The maximum distance of the NTF \filaa\ and \filab\ are
estimated to be 10.5 kpc from the Sun.

The present study extends the number of NTFs, which have
been found to be located near the GC region to five. With
most of the known NTFs now being shown near the GC, there
remains little doubt that phenomena related to the central
region of the Galaxy are responsible for the creation and
maintenance of the NTFs.

\begin{acknowledgements}
It is a pleasure to thank A. Pramesh Rao, with whom I have
discussed several aspects of this work at various stages. I
also thank Jayaram Chengalur, Miller Goss, Rajaram
Nityananda, Dharam Vir Lal and Nimisha Kantharia for reading
the manuscript and for their useful comments. Cornelia Lang,
as the referee, has made several useful comments, which
helped to improve the paper, and I would like to thank her.
Masato Tsuboi and Tomoharu Oka provided their CS and CO
survey data respectively and I wish to thank them. I thank
the staff of the GMRT that made these observations possible.
GMRT is run by the National Centre for Radio Astrophysics of
the Tata Institute of Fundamental Research.  I acknowledge
of receiving partial funding from the Rekhi Scholarship of
the TIFR Endowment Fund. 

\end{acknowledgements}
\bibliographystyle{aa}
\bibliography{roy.hi,aa}
\end{document}